\documentclass[]{article}

\usepackage{cite}

   \usepackage[pdftex]{graphicx}
  
\usepackage{amsmath}
\usepackage{breqn}

\usepackage{algorithmic}

\usepackage{array}

\usepackage{fixltx2e}

\usepackage{stfloats}

\usepackage{url}

\usepackage{multirow}
\usepackage{longtable}
\usepackage[export]{adjustbox}
\usepackage{tabularx,booktabs}
\newcolumntype{C}{>{\centering\arraybackslash}X}
\usepackage{caption}
\usepackage[dvipscolors]{xcolor}
\usepackage[colorinlistoftodos]{todonotes}
\usepackage{color}

\begin{document}

\title{Investigating Classification Techniques with Feature Selection For Intention Mining From Twitter Feed}

\author{Qadri Mishael and Aladdin Ayesh \\
	Faculty of Computing, Engineering and Media, \\ De Montfort University, Leicester LE1 9BH, UK\\
	Emails: \{qadri.mishael, aayesh\}@dmu.ac.uk 
}

\date{4 March 2019}
%
%

\maketitle

\begin{abstract}
	
In the last decade, social networks became most popular medium for communication and interaction. As an example, micro-blogging service Twitter has more than 200 million registered users who exchange more than 65 million posts per day. Users express their thoughts, ideas, and even their intentions through these tweets. Most of the tweets are written informally and often in slang language, that contains misspelt and abbreviated words. This paper investigates the problem of selecting features that affect extracting user's intention from Twitter feeds based on text mining techniques. It starts by presenting the method we used to construct our own dataset from extracted Twitter feeds. Following that, we present two techniques of feature selection followed by classification. In the first technique, we use Information Gain as a one-phase feature selection, followed by supervised classification algorithms. In the second technique, we use a hybrid approach based on forward feature selection algorithm in which two feature selection techniques employed followed by classification algorithms. We examine these two techniques with four classification algorithms. We evaluate them using our own dataset, and we critically review the results.


\end{abstract}

\textbf{Keywords} - Text Mining, Feature Selection, Intention Mining, Micro-blogging, Classification

\section{Introduction}\label{s-intro}

Online social networks have become an essential part of people's life nowadays. Social networks users like to share information by publishing posts about daily activities, feelings, opinions, interests or goals. The posts vary in content type to include text, images, video clips, or even URLs.  
Even though discovering human intention from the social networks posts depends on the reader understanding, researchers attempted to extract the intention. The researchers acknowledged intention detection within social networks as a valuable source of information to understand human behaviour and needs as in customers of online businesses \cite{Khan2016}.

The importance of studying human intention is acknowledged in various research disciplines such as psychology, sociology, and computer science. Consequently, there are several definitions for the intention was used. To give an example, \cite{Bratman1991} defined human or "an agent" intention as the state of mind at the time of taking action. In this paper, we adopt the following definition which was adapted from \cite{Purohit2015} : in a given system, user's intention is represented in the form of the user's goal of performing an action or a set of actions. A user's action could be surfing a web page online, publishing posts over a social network, or making an online query over a search engine. This research focuses on users’ intention in posting over social network platforms.  
Detecting the users' purposes and goals from their actions and instructions is known as Intent Mining \cite{Ding2015,Purohit2015}. The goal behind Intent Mining is to enhance the services that a system provides to users. Intent mining focuses on collecting and analysing the user's preferences and data from the system through systems logs \cite{Chen2002,Horvitz1998,Khodabandelou2015}, browsing history, web search data\cite{Chen2014,Zhang2016a,Vineet2014} , or online social data \cite{Salaheldeen2015,Chaouali2016,Banerjee2012,Purohit2015,Park2016}.\\
System logs hold all system-user interactions as clicks or browsing history. Web page queries and link access also provide rich information about users. In social networks, users publish their desires, wishes, likes, and dislikes and share it with others.
The intent can be either explicit or implicit for the readers. The intention can be expressed explicitly in the form of posts over social networks, or queries on web pages, etc. To give an example, a user searching for a restaurant in a particular area or posting "heading to city centre for lunch!", displays an explicit the intention for travelling somewhere, the "city centre", with a goal of having food.  On the other hand, some users’ actions indicate implicitly user intention, from the forehead example, the user is looking for restaurants to have his lunch even though the "restaurant" word is not mentioned in the post. Furthermore, extracting users' intent using a machine with an accurate understanding of the users' needs and goals from the system is a challenging process. For instance, the posts on social networks are not presented in a clear format for mining, as the language that used is usually informal, with abbreviations, misspelling, emot-icons, hash-tags, or even having multiple data formats as in images, audio, and videos. This paper is limited to social network data and microblogging social network posts in specific. 

In this paper, we study the feature space to determine and identify the features that define intention over the online social network. We applied text mining mechanisms to extract user's intention features from social media in general and microblogging in particular.  Knime as a data mining tool is used to implement feature selection and classification techniques. Section \ref{s-prel} discusses the preliminaries and the previous studies on intention mining and online social networks, followed by, description of the dataset, data mining tools, and data corpus and how to retrieve the social data online in section \ref{s-corp}. In addition, the applied schemes are described in section \ref{s-exp}, followed by, discussion of the experiments and results in section \ref{s-disc}. Finally, conclusion and future works are presented at the end of this paper.

\section{Preliminaries}\label{s-prel}

Analyzing social networks to extract patterns from users' data gives better understanding to the human behaviour in general and human intentions in specific. In this section, we look into intention mining literature, and review the social network microblogging.

\subsection{Intention Mining} \label{s-Back}

Intention mining has been an active research area in the recent years. A number of researchers showed interest in discovering individuals’ intention through studying the online behaviour and public data of those individuals. The users’ information is retrieved from web queries, systems logs, and social networks. In this paper, we are studying the features that define intention as a classification problem over social networks in general and feature selection in specific. Various data mining techniques and feature selection algorithms are found in the literature. The researchers in \cite{Dai2006} studied the Web pages textual features that resulted from users Web searches and queries to predict the users’ commercial intention for her online activities. Their prediction technique was based on the Support Vector Machine(SVM) algorithm. Chen et al., in \cite{Chen2013a}, worked on identifying users’ intents over forums posts based on Information Gain method given by \cite {Yang1997} for feature selection and Expectation Maximization algorithm. Vineet et al.\cite{Vineet2014}, also, worked on the linguistic features of the expressions over Yahoo!answers and Quora to extract user purchase intention. They used 'bag-of-words' to extract features. Moreover, Ding et al \cite{Ding2015} used word embedding feature selection technique as in Convolution Neural Network (CNN) to identify the consumption intent of users over micro-blogging in China. Their model of consumption intention was based on a binary classification of the posts to decide if the sentence contains consumption intention or not.
Furthermore, Kim et al. \cite{Kim2016} studied the travelling intention of the social network users. They built a textual features vector from the users shared information over social networks using word embedding techniques, and the researchers used classification algorithms such as Random Forest, SVM, and Deep Neural Networks(DNN) and Naïve Base(NB) to validate the created vector.  In Zhang et al.\cite{Zhang2016a}, applied neural network algorithms over text queries to capture user intention from online medical queries. Another approach, by \cite{Wang2015a}, proposed using intent keywords instead of using bag-of-words to apply a graph based semi-supervised learning technique for mining user intent and classify the tweets into six categories. From literature, it becomes obvious that the problem intention-mining of the users of the online social network, in general, depends on extracting the textual features that defined the user’s intention over the online social network, and these features are used in predictive models that built using the classification techniques.

\subsection{Microblogging} \label{s-micro}

Many Platforms with dedicated user interfaces are used to enable users to access their related social networks easily, by using different technologies as in mobile devices. Consequently, social network platforms provide us with input, e.g. Twitter feeds, for our intention mining as in the followed sections. Microblogging platforms distinguish from other social network platforms in having a short text messaging that avoid any information overload. In addition, it differs in making all the content publicly available to the other users in the platform. It also differs from the common blogging by having a limited number of characters per post. 

Taking \textbf{Twitter\footnote{https://www.twitter.com/}} for instance, it is one of the most well known micro-blogging services. users can follow other users in Twitter or can be followed with no need for any reciprocation. Twitter users get information about "what are you doing or thinking" as tweets of their Twitter friends in real time \cite{Kwak2010}. One of the differences between Facebook and Twitter is that, Facebook used to help users to interact and communicate with their friends and family in the real world, while Twitter helps users to communicate with any person who shares the same interest. 
Twitter has both website and a related mobile application, and associated APIs that support applications' programmers and enable them to develop new functions and services for the Twitter platform. Moreover, Twitter enables mobile device users to send new tweets to the Twitter web site not only through the mobile application, but also by short messaging service (SMS). 

\textbf{Tumblr\footnote{https://www.tumblr.com/}} is another example on micro-blogging platforms which popularity comes from storytelling using gif images that users add to their posts make it more descriptive. It also supports multimedia as audio and videos. Tumblr users can share external contents from other sources such as articles or external URLs by adding them to their posts \cite{Agarwal2017}. In this paper, our interest is in Twitter due to the wide availability and extensive use by English speaking users. In addition, its APIs are easy to use and supported by many data mining tools. There is a possibility to apply our further research on Tumblr in the future but not in this paper. 

As our current research is focused on the micro-blogging feeds in English, we limit our consideration to those services that are primarily used by English speakers. However, the growth of micro-blogging services in a wide range of languages, e.g. Sina Weibo \cite{Zhao2014}, makes the generalisation of our research reported here to other languages an interesting venture.

\subsection{Feature Selection} \label{ss-fs}

Feature selection is the process of selecting a proper minimum features set from the overall features available. This process is achieved by taking out any irrelevant features and remove any redundancy \cite{Xue2016}. This reduces the dimensionality of the data and increase the performance of executing the classification algorithms \cite{Xue2016}. Feature selection algorithms can be categorized into supervised, unsupervised, and semi-supervised feature selection. Supervised feature selection methods can be categorized into wrapper models, filter models, and embedded models and hybrid models. The wrapper-based models  generate the subsets of features using any one of the searching techniques and evaluates these subsets using the  supervised classification algorithm in terms of accuracy. Wrapper method has some noticeable defects such as searching overhead, over-fitting, and increased runtime. The  embedded based feature selection models use a part of the learning process of the supervised learning   algorithm  for feature selection. Embedded models offer less computational cost comparing to the wrapper models. Yet, embedded models suffer from poor generality. This  embedded  models are categorized into three namely pruning method, built-in mechanism, and  regularization   models. The filter feature selection models are independent of the supervised learning algorithm therefore consider more general and computationally cheaper comparing to  the wrapper and the embedded approaches. Therefore, filter models are better in processing high-dimensional data rather than the wrapper and embedded methods \cite{AsirAntonyGnanaSingh2016}. Among the most representative algorithms of the filter model we have: Relief, Fisher score, Information Gain. The hybrid  feature selection models are based on the combination of different approaches as filter and wrapper-based approach.The feature selection is considered an initial step in supervised data mining analysis, yet it is a challenging problem, especially for social post that are massive, noisy, and sometimes incomplete \cite{Tang2014}. There is a need for feature selection algorithm that can deal with such data. In social networks, the elements with high-dimensional features are often linked together. Another difficult problem is how to integrate link information to guide feature selection \cite{Li2016a}. In this paper, we applied hybrid approach to by using Information gain as filter feature selection model and wrapper approach since the problem is supervised learning problem.

\section{Social Data Corpus}\label{s-corp}

The focus of our research is the social network’s feeds; therefore, we need to have a dataset that hold enough information. The needed dataset considered as Big-data that need to be analysed using special tools. The tools should have different techniques for extracting the needed information in order to achieve our goal. Examples of the datasets that have been used in the literature are revised in this section. Similarly, key data mining tools are discussed. 

\subsection{Datasets}\label{ss-datse}

Twitter has been used as target of this paper. Several online datasets that are published by researchers that focus on the sentiment analysis but not much related to intention mining.\cite{Go2009a,Mohammad2017,SemEval2018,Saif2018} 
For example, Sentiment 140 \footnote{http://www.sentiment140.com/} dataset which is available for research purposes containing \textbf{1600000} records for training and \textbf{497} records for testing. The data was collected from Twitter in 2009 \cite{Go2009a}. The dataset was used in  WASSA shared task on emotion (EmoInt) and included in the \textit{AffectiveTweets}\footnote{https://affectivetweets.cms.waikato.ac.nz/} a Weka package for analyzing emotion and sentiment of English written tweets\cite{Mohammad2017}. 


SemEval datasets are well-known datasets for carrying out number of semantic analysis  tasks for text in a series called SemEval(Semantic Evolution). In the time of writing this paper the SemEval-2018 task 1 dataset was published for affect in tweets \cite{SemEval2018}. This dataset consists of 100 millions tweets ids. The tweets collected based on emotion-related words such as angry, annoyed, panic, happy, elated, surprised, etc\cite{Saif2018}. However, the online datasets that used Twitter usually published in anonymised form. Indexes are used to hide the tweets and users, and that make the reuse of these datasets a complicated process. A reverse engineering process is needed in order to retrieve the original tweets which takes time and efforts, and in many cases the tweets could not be founded since they are deleted by the users. In our work, the original posts are required to be analysed for intention and processed for feature extraction. In addition, we intended to conduct our experiments in a control environment. Therefore, we crawled Twitter using Application Provider Interfaces (APIs) to build our own dataset. 


\subsection{Data Mining and Analysis Tools}\label{ss-dmto}

Data mining tools such as Knime, Weka, Orange and RapidMiner, etc are used by researchers to study and analyze structured and unstructured data collections. For our work, we explored these state of art tools and evaluated them in relation to their GUI, ease of use, and supported algorithms.
 
\textbf{Knime} \footnote{https://www.knime.org} (Konstanz Information Miner) is an open source workflow data mining platform based on Java and Eclipse platform. It works under different operating systems Windows, Linux, macOS \cite{Rangra2014}. In addition, it supports big data analysis with graphical user interface. Its visual interface gives the ability to access data and apply data transformation and it supports powerful predictive analytics \cite{Gibert2012}. Knime workflow consists of connected nodes or extensions\cite{Berthold2013}. Moreover, Knime supports integration of different data analytic tools such as R, Python scripting, Weka, and other third party applications such as Google Analytics. Furthermore, Knime provides nodes for connection to social media platform such as Twitter. This integration makes the use of Knime suitable for our work. Each Node in Knime takes a part in processing data before passing it to the following nodes through their connections. The data are stored in each node in a table format. The tables could be saved permanently at any point to be processed in a different format. Due to the expandability of Knime, new nodes can be added at any point to apply different kind of processing without the need to re-execute the previous nodes. Knime can be downloaded and used freely under an open source license (GPL) \cite{Berthold2013}.  

\textbf{Weka} \footnote{https://www.cs.waikato.ac.nz/ml/weka/downloading.html} stands for (Waikato Environment for Knowledge Analysis), which is a free data mining tool based on Java. It is supported from different operating systems. It combines several tools of data preprocessing, machine learning, visualization, and feature selection. Weka user interface built of several components, which are accessed through an Explorer. In addition to access component-based knowledge flow interface and the command line. One of Weka's features is the Experimenter component, which facilitates executing a systematic comparison on a collection of datasets and applying several machine-learning algorithms at once. Weka has GNU general public license, which make it free to install. However, Weka does not support connecting to non-Java based databases, and sometimes fails in reading CSV files \cite{Rangra2014}. 

\textbf{Orange} \footnote{https://orange.biolab.si} is a data-mining tool that implemented in C++ and Python. This tool supports different operating systems. It has different data mining and machine learning algorithms. Python libraries should be installed in order to have this tool running smoothly. Different components are provided for data preprocessing, feature filtering, data modelling and evaluation, and visualisation. However, it has a limited reporting capability. 

\textbf{RapidMiner} \footnote{https://rapidminer.com} is a stand-alone application with user friendly interface that supports various operating systems. It works as an integrated platform and supports different machine learning and data mining techniques including text mining, with predictive analytics and business analytics. It adopts graphical ETL (Extraction-Transform-Load) process workflows. The simplicity use of Rapid Miner Studio can be seen in the drag and drop operations for the operators, setting parameters and combining operators. Moreover, it supports different data input and output file formats and can be connected to relational and non-relational databases. Yet, Rapid miner requires a knowledge of SQL and database handling. In this paper, we implemented our work using Knime due to the availability of the aforementioned characteristics.

Table \ref{t-Tools} summarises our investigation in relation to the following factors: GUI,	Ease of Use (EoU), Connectivity (C), supported Data Processing Algorithms (DPA), Machine Learning (ML) techniques available, the availability of Data Visualisation (DV), and Programming Languages (PL) supported. 

\begin{table*}

	\caption{Tools Analysis}
	\label{t-Tools}
	\centering
\begin{tabularx}{1.2\textwidth}{@{}l*{14}{C}c@{}}
\toprule
		Tool	& GUI &	EoU &	C  &	DPA   &	 ML   &	 DV  &	PL \\
		\midrule
		Knime & Windows, Mac, and Linux & Easy (drag and drop) & relational database, NoSQL databases, and cloud services & Supported & Supported & Supported &	Java, Python, R						\\
		Orange & Windows, Mac, and Linux & Easy (drag and drop) & Orange delimited format, SQL & Supported & Supported & Supported &	Python					\\							
		WEKA	& Windows, Mac, and Linux & Drop down lists & Weka special formats and Java based databases   & Supported & Supported & Supported & Java					\\						
		RapidMiner	& Windows, Mac, and Linux & Easy (drag and drop)& relational database, NoSQL databases, and cloud services & Supported & Supported & Supported &	R and Groovy					\\						
\bottomrule		
\end{tabularx}
\end{table*}

\subsection{Building Social Corpus From Twitter Post}\label{s-bsc}

Since we could not find an intention corpus for twitter post that serve our research, we built our own corpus SICorp and adopted the following formal representation of corpus intention classes:

Suppose a corpus $R$ of $n$ short statements documents set $D$ , where $D = \{d_i | 1\leq i \leq n\}$, and a set $C$ of predefined $m$ target classes,
\begin{equation} 
	C = \{c_j| 1\leq j\leq m\}
\end{equation} 
There will be a prediction function $f$ such that
\begin{equation}
	f = \{D \times C\}\Longrightarrow \{YES , NO\}
\end{equation}
	which is indicate that an intention $c_j$ in the intention set $C$ is presented in the record $r_i$ in the corpus $R$. 
\begin{equation}
	\exists c_j \in C \Longrightarrow r_i \in R 
\end{equation}
The class set for the research is a text vector: \\
$V_w=\{wish, want, need, look for, request, like, desire \}$. \\
These words were selected as the initial seeds for retrieving social post. 

Twitter APIs make collecting large number of tweets a relatively easy task since it supports different programming languages and data mining tools. In order to retrieve published posts, a certain condition or conditions should be set, such as term or terms included in the post, the user who created the post, the location of the user, or the language of the retrieved posts. In addition to the retrieved post, Twitter APIs return information such as the tweet id, publication date and time, author's username and id, location, hashtags, number of retweets, number of followers and friends, the language and other data.

Twitter API was used through Knime tools to connect to Twitter and used to collect data; the connection requires API Key and Access Token. Certain search queries are used to retrieve the tweets.
Our dataset is built with certain conditions. First, we retrieved all the tweets that contain any of the words  that presented in the words vector with total number of 7000 tweets. Second, we filtered the tweets to be limited to the English ones, also, we removed all the advertisements posts. The number of tweets was reduced to 5896 in the English language. The number of tweets consider sufficient to conduct our exterminates on a small scale.    

We assumed that each post represent a single sentence since the post cannot exceed 140 characters by Twitter platform rule. Based on this assumption, we considered the intention within the post to represent user's intention and all the words of the post are leading to this intention. The posts are represented in a text format and labelled based on the intention word that introduced in the search vector. The algorithm divided posts into polarity class set as a target class set ${Yes, No} $, where \textbf{$Yes$} indicates the text contains intention words and \textbf{$No$} is the opposite. The class distribution over the dataset which is 3452 tweets labelled as \textbf{$Yes$} and 2444 \textbf{$No$} labelled tweets with difference of 17\% for the \textbf{$Yes$} class. 

\section{Feature Selection Scheme for intention mining}\label{s-exp}

Two schemes were conducted Schem1 and Schem2, each had two parts . The first part in both schemes is the selection of features from the dataset.  
 In  Schem1, we used machine-learning Information Gain (IG) algorithm feature selection. In Schem2, machine-learning algorithms are used back to back as a Hybrid feature selection. Both are followed by a supervised classification algorithm to classify the dataset based on the selected features using four machine-learning algorithms, which are Decision Tree, Naive Bayes, Support Vector Machine, and Feed-Forward Learner Neural Network.
  The accuracy of those classification algorithms reflects the quality of the feature selection techniques.

\subsection{IG Feature Selection} \label{ss-exp1} 

In this part of scheme1, we focused on extracting the features of the posts that specify intention, and selecting the features by applying Information Gain (IG) Algorithm. IG is used as a machine learning technique to predict features according to the terms in text \cite{Yang1997}. It measures the number of text features obtained for the category prediction by knowing the presences or absence of a term in a text. IG value for term $t$ calculated as follows:
\begin{dmath}
	IG(t) = - \sum_{t=1}^{m} P(c_i)log P(c_i) + P(t)\sum_{t=1}^{m}P(c_i|t)log P(c_i|t)+P(\overline{t})\sum_{t=1}^{m}P(c_i|\overline{t})log P(c_i|\overline{t})
\end{dmath}
where $c_i$ stands for the set of the categories in the target space; $P(c_i)$ stands for the probability of category occurs; $P(t)$ is the probability of term $t$ occur;$P(\overline{t})$ represent the term $t$ does not occur; where $m$ is the total number of target classes.  
IG usage reduces the dimension of the features and speed up the classification processes. Tweet vector is used to select the set of features that will be used for classifying tweets into two classes, \textbf{$Yes$} as if an intention exists in the text and \textbf{$No$} if there is no intention. 

A Tweet Vector ($VT$) represents a word vector of terms in tweets space as binary values. The extracted feature vectors is constructed using  \textbf{ Bag of Words model (BOW)} and \textbf{Term of Frequency (TF)}. BOW creates a vector of unigrams for the terms that exist in the text based on PoS tagging that done the preprocessing phase \cite{Rout2018}.
TF is used to calculate the frequency of the term in the text, terms considered as features to be extracted \cite{Yang1997, Baccianella2013}.

\subsection{Hybrid Feature Selection}\label{ss-exp2}
In Scheme2, a threshold of the features is set to specify the number of features that reach maximize score in the form of terms vector. The scheme is built into two parts. The first part is hybrid of feature selection based on two different algorithms. Starting by selecting the features based on IG Algorithm. 
IG algorithm extracted eighty two features as in Scheme1. The second phase of feature selection is applied using Forward-Feature Selection algorithm (FFS). FFS starts by building an empty set of features and adds one feature at time to the set and start evaluating.
The algorithm depends on measuring the Leave-One-Out Cross Validation (LOOCV) error of the one-feature subset to find the  best individual feature.
Four algorithms FFS are applied to select features that are NB, SVM, ANN, and DT. FFS feature threshold is set on least number of features that give the maximum accuracy score which is in our case was different with each algorithm applied. Those features are expected to be the words that used in the Twitter feed. In the second part, four classification algorithms are used as DT, SVM, ANN, and NB to test the quality of the selection technique that used.\\

\section{Experimental Results} \label{s-disc}

\begin{figure}[h!]
\captionsetup{justification=centering}
\includegraphics[width=0.9\linewidth]{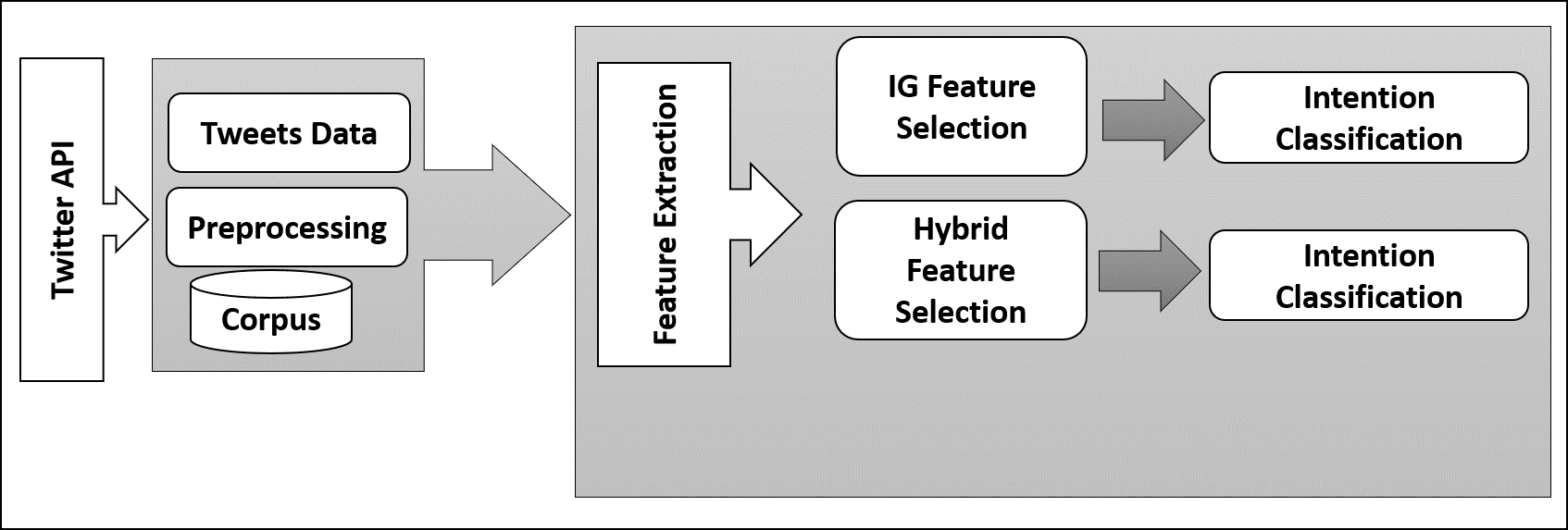}
\caption{Experiments Framework}
\label{fig:framework}
 \end{figure}

The experiments have been carried out based on the scheme described in section \ref{s-exp}. The experiments were designed to test the possibility of mining the users' intention by applying data mining techniques on the SICorp corpus.
 This section describes the setup of the experiments, followed by explaining how the data has been preprocessed due to the informality and noisy nature of the social posts. Furthermore, we analyse the results for each algorithm that has been used. We conclude with a critical analysis of the experiments results. We look at the feature selection using IG experiment that followed by one of the classification algorithms DT, ANN, NB, and SVM. Following sections, we look deeply in the performance of the classification algorithms (DT, ANN, NB, and SVM) after conducting the second experiment (the hybrid feature selection that used IG+DT, IG+ANN, IG+NB, and IG+SVM). Figure \ref{fig:framework} illustrates the experiments framework.

\subsection{Experiments Settings}\label{ss-expset} 

The Knime 3.4 64-bit platform is used through all experiments' phases. The used machine is operated by Windows 10 64-bit operating system. The machine processing power and memory are Intel i7 2.2GHz and 8 GB - 12 GB RAM respectively. The dataset is described in section \ref{s-corp}-\ref{ss-datse}. 
We start our experiments using 8 GB RAM machine. However, we faced many difficulties in running the experiments such as long execution time and OutOfMemory problems. These problems noticed when executing the machine learning algorithms for extracting features from 2836 features, consuming a massive portion of memory 6 GB - 8 GB which lead to OutOfMemory problem. In addition, the same problem occurred when we used the whole set of features 2836 for classification part of the experiment, specially with SVM algorithm due to its computational complexity. increase the machine memory to 12 GB. 

\subsection{Preprocessing Data} \label{sss-predata}

The collected dataset from Twitter is noisy and difficult to analyse due to language informality, misspelling, emot-icons, and URLs, therefore, several preprocessing steps are considered. 
These preprocessing steps are presented using text mining techniques. In the first preprocessing step, (\textbf{Preprocessing-1}), 
all the URLs are removed from the text to eliminate any conflicts of having URLs. The URLs could be fall words with weight when building the word vectors on the following steps. 
In the second preprocessing, (\textbf{Preprocessing-2}), a set of different text filtering techniques is employed. These techniques are; Part of speech (PoS) tagging which used to tag each term based on its position in the sentence a noun, an adjective, verb, or adverb. Also, to remove all the punctuations symbols within the text and the stop words. The output of this step is filtered text of each tweet record prepared to apply feature extraction and classification algorithms. 

\subsection{Experiment 1: IG Feature Selection based Classification}

Before applying the feature selection techniques we trained four well-known classification models, which are Decision Tree Classifier (DT), Support Vector Machine (SVM), Feedforward Learner Neural Network (ANN), and Naive Bayes (NB) for the whole  sets of labelled tweets using all 2836 features, the results are shown in table \ref{table0}. The goal of this step is to setup a benchmark to test the impact of feature selection on the classifiers performance. As noticed from the table\ref{table0}, when all the features are provided the highest measure in (F-measure) and accuracy metrics for training ANN as 86.73\% and 84.07\% respectively. While the NB produce the worst performance as 61.79\% and 72.55\% for both accuracy and (F-measure) metrics.

\begin{table}[h!]
	\caption{Experiment result of using all features before applying the feature selection technique by a classifier algorithm, which are  NB-Naive Bayes, SVM-Support Vector Machine, and ANN- Neural Network}
	\label{table0}
	\centering
	\begin{tabular}{@{}ccccc@{}}
		\toprule
		\textbf{Classification} & \textbf{Recall}  & \textbf{Precision} & \textbf{F-measure} & \textbf{Accuracy} \\ \midrule
		\textbf{DT}              & 87.77\%          & 83.17\%             & 85.41\%             & 82.51\%             \\ 
		\midrule
		\textbf{ANN}             & 89.25\%          & 84.34\%             & 86.73\%             & 84.07\%             \\ 
		\midrule
		\textbf{SVM}             & 99.64\%          & ~66.05\%            & ~79.44\%            & ~69.94\%            \\ 
		\midrule
		\textbf{NB}              & 86.64\%          & 62.40\%             & ~72.55\%            & ~61.79\%            \\
		\bottomrule
	\end{tabular}
\end{table}

\begin{figure}[h!]
	\captionsetup{justification=centering}
	\includegraphics[width=0.7\linewidth]{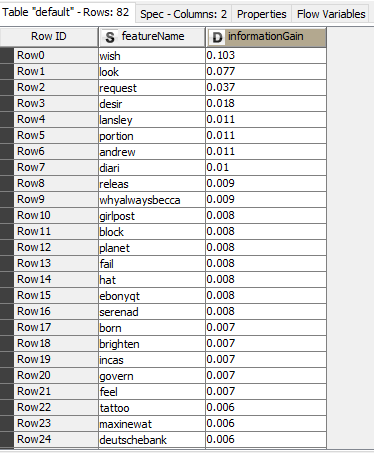}
	\caption{ The Output of Applying Information Gain}
	\label{fig:IGresult}
\end{figure}

Information Gain (IG) uses the entropy to measure the uncertainty between text and target class with and without the features. This means the most important features to classify the tweets are used. It is widely used to extract features from text \cite{Liu2017}.
In the first scheme1 , IG feature reduction technique was applied which selected eighty-two features from 2836 features on the whole sets of collected tweets, see figure \ref{fig:IGresult}. 
Because IG algorithm calculates the mutual information ratio of the dataset, the selected features have the highest mutual information ratio, i.e. all the eighty-two terms information gain ratio is greater than zero $(IG(t) > 0)$ while the rest of the 2754 features is equal to zero. 

The eighty-two features are used to train classification models (DT), (SVM), (ANN) and (NB)for the labelled tweets. In the classification phase, with ten folds Cross-Validation setup of leave one out. Hence, for each fold of the cross-validation the algorithm is trained on all the items except one-instance. Although, feature extraction was not based on the context of the tweets, IG algorithm reduced the features significantly.
In table \ref{table1}, it is noticed that applying IG by itself to extract features provides the highest measure in (F-measure) metric for training DT as 86.14\%. This is considered as the selected features hold enough information to give prediction. 

\begin{figure}[h!]
	\centering
	\captionsetup{justification=centering}
	\includegraphics[width=0.9\linewidth]{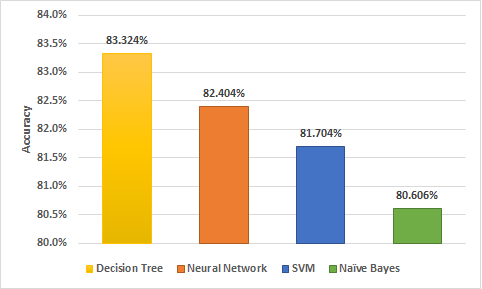}
	\caption{The accuracy of applying Information Gain feature selection in Experiment 1}
	\label{fig:IGperformance}
\end{figure}

The illustration of the accuracy for the learning classification models over the collected dataset is presented in figure \ref{fig:IGperformance}. In the table the  

\begin{table}[h!]
	\caption{Experiment result of using 82 features selected by Information Gain as a feature selection technique followed by a classifier algorithm, which are  NB-Naive Bayes, SVM-Support Vector Machine, and ANN- Neural Network}
	\label{table1}
	\centering
	\begin{tabular}{@{}ccccc@{}}
		\toprule
		\textbf{Classification} & \textbf{Recall}  & \textbf{Precision} & \textbf{F-measure} & \textbf{Accuracy} \\ \midrule
		\textbf{DT}	& 88.86\%	& 83.58\%   & 86.14\%   & 83.32\% \\ \midrule
		\textbf{ANN}	& 88.36\%	& 82.66\%	& 85.41\%	& 82.40\% \\ \midrule
		\textbf{SVM}& 84.30\%	& 82.75\%	& 83.52\%	& 81.70\% \\ \midrule
		\textbf{NB}	& 85.74\%	& 83.35\%	& 84.53\%	& 80.61\% \\ \bottomrule
	\end{tabular}
\end{table}

\subsection{Experiment 2: Hybrid Feature Selection based Classification}
\subsubsection{Decision Tree (DT)}

DT setup, which is based on C4.5 \cite{Shafer1996}, for the experimentation was as follows: 
\begin{itemize}
	\item The quality measure is calculated based on the Gini Index as splitting technique, with no pruning \cite{Shafer1996}. 
	\item The minimum number of nodes is 2. The split point value is calculated according to the mean value of the two attribute values that separate two partitions. Working on eight cores to speed performance.
\end{itemize}

Since the decision trees learning method predicts the values of target variable by learning simple decision rules inferred from the data features, it resulted in a relative high outcome. It is robust to noisy data, and since it is a heuristic algorithm, that means a decision is obtained locally and does not guarantee to return the globally optimal solution.
\begin{table}[h!]
	\caption{Experiment results of using Decision Tree classifier with eight features selected through the Hybrid feature selection}
	\label{t-DT}
	\centering
	\begin{tabular}{@{}ccccc@{}}
		\toprule
		\textbf{Feature Selection}	& \textbf{Recall} & \textbf{Precision} & \textbf{F-measure} & \textbf{Accuracy} \\ \midrule
		\textbf{IG+NB}	&  88.56\%	& 83.61\%	& 86.02\%	& 83.21\%  \\ \midrule
		\textbf{IG+SVM}	&  88.56\% 	& 83.61\%	& 86.02\%	& 83.21\%  \\ \midrule
		\textbf{IG+ANN}  &  88.76\% 	& 83.64\%	& 86.12\%   & 83.32\%  \\ \midrule
		\textbf{IG+DT}  &  88.46\%	& 82.44\%	& 85.35\%   & 82.29\%  \\  \bottomrule
		
	\end{tabular}
\end{table}

Applying two phases to reduce feature gives a relativity close results, even though, the features are reduced to eight. The reduction of the features reduces data processing time, yet, the accuracy is slightly less. By observing Table \ref{t-DT}, almost the same accuracy values have been resulted for DT with very slight difference. 

\subsubsection{Naive Bayes (NB)}

The basic NB classifier is used to decide the right class of the input data by referring to the highest probability values that calculated by the trainer classifier using the Bayes formula. The right class is represented by the class which has the highest probability value as Bayes classification rule states \cite{Chen2009b,Lee2010}. The class is calculated as follows:
\begin{equation}
	P(c_j|d_i) = \frac{P(d_i|c_j).P(c_j)}{P(d_i)}
\end{equation}
Where $P(d_i)$ is the same for all the classes.
For applying the NB for feature selection in the second experiment, the probability of the word feature $w_k$ occurrence in a text document is independent of the word's position and the occurrence of other words in the text document. So the probability of $P(c_j|d_i)$ would be :
\begin{equation}
	P(c_j|d_i) =P(|d_i|)|d_i|!\prod_{k=1}^{|v|} \frac{P(w_k|c_j)}{n_ik !}^{n_ik}
\end{equation} 
Where $n_ik$ is the number of time that a word occurs in a document; and $|d_i|$ number of the words in a document.

Applying NB classifier with eighty two features produced the lowest accuracy in the IG feature selection from the first experiment with an accuracy of 80.61\% as shown in Table \ref{table1}. However, the accuracy increased when the feature set reduced to eight to eleven features in the second experiment as shown in Table \ref{t-NB}. 


\begin{table}[h!]
	\caption{Experiment results of using Naive Bayes classifier with the features selected through the Hybrid feature selection}
	\label{t-NB}
	\centering
	\begin{tabular}{@{}ccccc@{}}
		\toprule
		\textbf{Feature Selection}        &  \textbf{Recall} & \textbf{Precision} & \textbf{F-measure} & \textbf{Accuracy} \\ \midrule
		\textbf{IG+NB}	&  87.26\%	& 83.38\%   & 85.28\%	& 82.43\% \\ \midrule
		\textbf{IG+SVM}	&  87.26\%  & 83.38\%   & 85.28\%	& 82.43\% \\ \midrule
		\textbf{IG+ANN}  &  87.17\%  & 83.29\%   & 85.19\%   & 82.33\% \\ \midrule
		\textbf{IG+DT}  &  87.00\%  & 81.93\%	& 84.38\%	& 81.23\% \\ \bottomrule
	\end{tabular}
\end{table}

\subsubsection{Artificial Neural Network (ANN)}

The ANN algorithm is used based on FeedForward Learning with two inner layers with 100 output units each, and learning rate of 0.1. XAVIER initialization weight strategy \cite{Xavier2010} is used with ReLU Activation Function. The number of training iteration is one. The optimization Algorithm used is Stochastic Gradient Descent(SGD).  The loss function that used is Mean Squared Error.
Applying the hybrid feature selection shows an improvement as ANN improves the accuracy into 82.23\%, as seen in table \ref{t-ANN}. 
\begin{table}[h!]
	\caption{Experiment results of using FeedForward Neural Network classifier with the features selected through the Hybrid feature selection in experiment 2}
	\label{t-ANN}
	\centering
	\begin{tabular}{@{}ccccc@{}}
		\toprule
		\textbf{Feature Selection}        &  \textbf{Recall} & \textbf{Precision} & \textbf{F-measure} & \textbf{Accuracy} \\ \midrule
		\textbf{IG+NB} &  87.48\%	& 82.21\%	& 84.76\%	& 81.65\%  \\ \midrule
		\textbf{IG+SVM} &  88.17\%  & 82.55\%   & 85.26\%	& 82.23\% \\ \midrule
		\textbf{IG+ANN} &  87.97\%	& 82.29\%	& 85.0\%	& 81.94\%  \\ \midrule
		\textbf{IG+DT} &  88.36\%	& 81.83\%	& 84.97\%	& 81.77\% \\  \bottomrule
	\end{tabular}
\end{table}

\subsubsection{Suport Vector Machine (SVM)}

SVM which is based on LibSVM algorithm \cite{Chang2013} has been used with overlapping penalty set to one, kernel used is Radial Basis Function (RBF) with Gamma equals to one.
For SVM classification technique, the highest accuracy is reduced when applying hybrid comparing to IG as shown in tables \ref{table1} and \ref{t-SVM}.

\begin{table}[h!]
	\caption{Experiment results of using Support Vector Machine classifier on the features selected through the Hybrid feature selection in experiment 2}
	\label{t-SVM}
	\centering
	\begin{tabular}{@{}ccccc@{}}
		\toprule
		\textbf{Feature Selection}        &  \textbf{Recall} & \textbf{Precision} & \textbf{F-measure} & \textbf{Accuracy} \\ \midrule
		\textbf{IG+NB}	&  86.10\%	& 82.55\%	& 84.28\%	& 81.28\%   \\ \midrule
		\textbf{IG+SVM}	&  86.10\%	& 82.55\%	& 84.28\%	& 81.28\% \\ \midrule
		\textbf{IG+ANN}  &  86.28\%	& 82.72\%	& 84.46\%	& 81.49\%  \\ \midrule
		\textbf{IG+DT}  &  85.83\%	& 81.45\%	& 83.58\%	& 80.34\%  \\  \bottomrule
		
	\end{tabular}
\end{table}

\subsection{Critical Analysis}\label{ss-resana}

The two experiments that aimed to predict the existence of intentions on the feeds that users post on social networks. The prediction techniques were based on text features.

In figure \ref{fig:IGperformance}, the accuracy of the classification learning models is illustrated over the collected dataset, using IG algorithm as a feature selection technique produce higher accuracy for feature selection comparing to other techniques that are used in the second the experiment. The DT C4.5 produced the highest accuracy locally comparing to the other classification techniques followed by ANN, SVM, and NB.

In the second experiment, applying a hybrid feature selection to reduce the number of features did not show a significant difference in classification results. This means that with a minimum number of features it is possible to get the very close accuracy as using eighty-two features. The features were reduced significantly from eighty-two to up to eleven features. 

By observing the tables \ref{table1}, \ref{t-DT}, \ref{t-NB}, \ref{t-ANN}, and \ref{t-SVM}, a slight difference in accuracy values between the four different classifier algorithms is noticed from applying the different techniques of feature selection. 
Using IG by itself with all the features or combine it with Forward ANN with the reduced features did not show any difference in the accuracy for the DT classifier as shown in table\ref{table1} and table \ref{t-ANN}.
Moreover, applying NB and SVM as a second stage feature selection resulted in same accuracy and F-measure for DT classification, as shown in tables \ref{t-DT}. Whereas, selecting ANN for the second stage showed a slight improvement in accuracy.
Since the DT learning method predicts the values of target variable by learning simple decision rules inferred from the data features, it resulted in relative high outcome.

\begin{figure}[h!]
	\captionsetup{justification=centering}
	\includegraphics[width=0.9\linewidth]{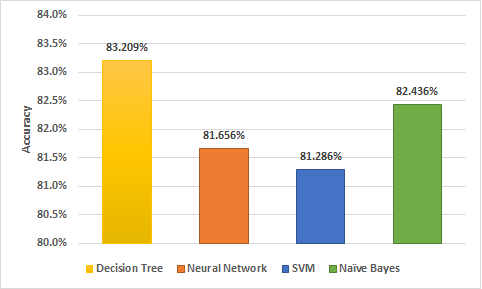}
	\caption{The accuracy of the four classifier algorithms after using IG and NB feature selection}
	\label{fig:subimg2}
\end{figure}
\begin{figure}[h!]
	\captionsetup{justification=centering}
	\includegraphics[width=0.9\linewidth]{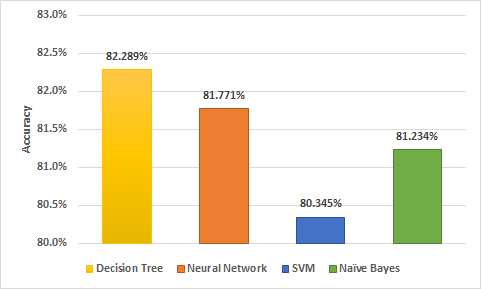}
	\caption{The accuracy of the four classifier algorithms after using IG and DT feature selection}
	\label{fig:subimg3}
\end{figure}
\begin{figure}[h!]
	\captionsetup{justification=centering}
	\includegraphics[width=0.9\linewidth]{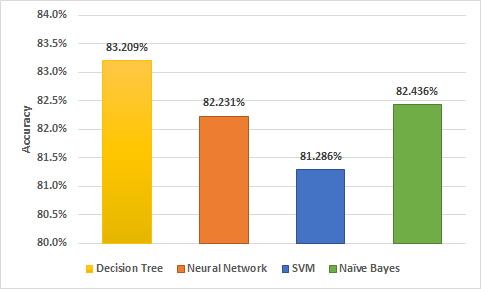}
	\caption{The accuracy of the four classifier algorithms after using IG and SVM feature selection}
	\label{fig:subimg4}
\end{figure}
\begin{figure}[h!]
	\captionsetup{justification=centering}
	\includegraphics[width=0.9\linewidth]{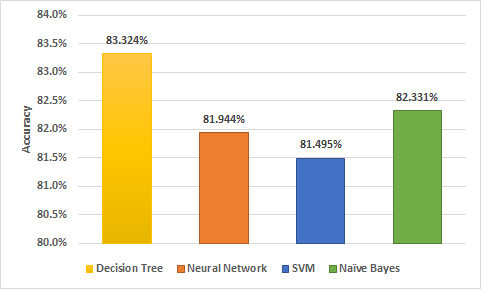}
	\caption{The accuracy of the four classifier algorithms after using IG and ANN feature selection}
	\label{fig:subimg5}
\end{figure}

NB classifier showed an improvement from adding another feature selection stage and reducing the features as shown in table \ref{t-NB}. In table \ref{table1}, the accuracy of using NB classifier after applying IG as a single feature selection technique was 80.61\%. Whereas , this accuracy increased to 82.43\% from using the hybrid feature selection in the second experiment as table \ref{t-NB} illustrates. 

Table \ref{t-SVM} shows that adding the ANN as second stage feature extraction gives a slight improvement in accuracy for applying SVM classification. However, applying the NB and SVM gives the same accuracy and F-measure but not for DT as a second stage.  
In addition, the accuracy from using the ANN as learner classifier improved when the SVM is added to the feature selection procedure. Table \ref{t-ANN}, illustrates the measurements for using ANN classifier. Reducing the features gave an advantage in increasing the speed of processing the data.

From figure \ref{fig:subimg2}, we can conclude that NB classifier can be improved when having the two phase feature selection especially as NB, SVM, and ANN. The ANN classifier improved when adding a DT as a second phase of feature selection. Whereas, SVM showed better performance when adding the ANN to the feature selection technique.

However, The eighty-two features that were extracted using IG are not considered based on the context, which can be understood by a human reader. As most of the output results are considered high, still it does not reflect the context of the tweets in a way that represents all features of social users’ intention. Some factors have to be considered in the future for this work, such as the accuracy of labelling. Labelling phase was done through applying search for a certain string within the retrieved tweets text, in other words, by labelling any tweets that have phrases from the intention vector, see section \ref{s-corp}, which is considered to hold an intention. Most of the previous studies made use of human-judge-labelling. Another factor is including more search words to retrieve the data from Twitter. More words patterns and terms are needed to be taken into consideration. Therefore, more experiments are needed to study the effect of applying different features from social network.

\section{Conclusion}\label{s-con}

Social networks have gained great interest from researchers because it provides a mean to study the human behaviour from online daily activities. There have been number of studies that focused on detecting intention of computer systems users. 
In this paper, we have looked at some datasets that are available online and used by other researchers. However, we faced difficulty in dealing with this datasets, therefore, we worked on extracting our own dataset from Twitter as a microblogging example. Different data mining tools have been reviewed here, and each one has its advantage and disadvantage. We used Knime tool because of its implementation of text mining techniques and the ability to save the result in different formats. In addition, it supports different programming languages that can help in implementing our model. The dataset was preprocessed using text mining filter techniques to remove any unneeded data such as URLs or symbols. The resulted dataset then used in two experiments. 
Both experiments had two parts, namely feature selection and classification. The first experiment had one feature selection phase using Information Gain. The second had two phases feature selection as a hybrid using Information Gain with three other algorithms. In both experiments, feature selection was used in classification and the performance of the algorithms was critically reviewed.


\def\url#1{}
\bibliographystyle{IEEEtran}

\bibliography{Ref}

\vfill


\end{document}